\documentclass[a4paper]{jpconf}
\usepackage{graphicx}
\usepackage{lineno}
\begin{document}
\title{27-day variation of the GCR intensity based on corrected and uncorrected for geomagnetic disturbances data of neutron monitors}

\author{M.V. Alania$^{1}$, R. Modzelewska$^{1}$, A. Wawrzynczak$^{2}$, V.E. Sdobnov$^{3}$, M.V Kravtsova$^{3}$}

\address{$^{1}$Institute of Mathematics and Physics, Siedlce University, Poland \\ $^{2}$Institute of Computer Sciences, Siedlce University, Poland,\\$^{3}$ The Institute of Solar- Terrestrial Physics of Siberian Branch of RAS, P.O.Box 291, Irkutsk, Russia.}

\ead{alania@uph.edu.pl, renatam@uph.edu.pl, awawrzynczak@uph.edu.pl, sdobnov@iszf.irk.ru}
\begin{abstract}
We study  27-day variations of the galactic cosmic ray (GCR) intensity for 2005-2008 period of the solar cycle {\#}$23$. We use  neutron monitors (NMs) data corrected and uncorrected for geomagnetic disturbances. Besides the limited time intervals when the 27-day variations are clearly established, always exist some feeble 27-day variations in the GCR intensity related to the constantly present weak heliolongitudinal asymmetry in the heliosphere.\\
We calculate the amplitudes of the 27-day variation of the GCR intensity based on the NMs data corrected and uncorrected for geomagnetic disturbances. We show that these amplitudes do not differ for NMs with cut-off rigidities smaller than 4-5 GV  comparing with NMs of higher cut-off rigidities. Rigidity spectrum of the 27-day variation of the GCR intensity found in the uncorrected data is soft while it is hard in the case of the corrected data. For both cases exists definite tendency of  softening the temporal changes of the 27-day variation's rigidity spectrum in period of 2005 to 2008 approaching the minimum of solar activity. We believe  that a study of the 27-day variation of the GCR intensity based on the data uncorrected for geomagnetic disturbances should be carried out by NMs with cut-off rigidities  smaller than 4-5 GV.

\end{abstract}
\begin{figure}[ht]
  \begin{center}
\includegraphics[width=0.7\hsize]{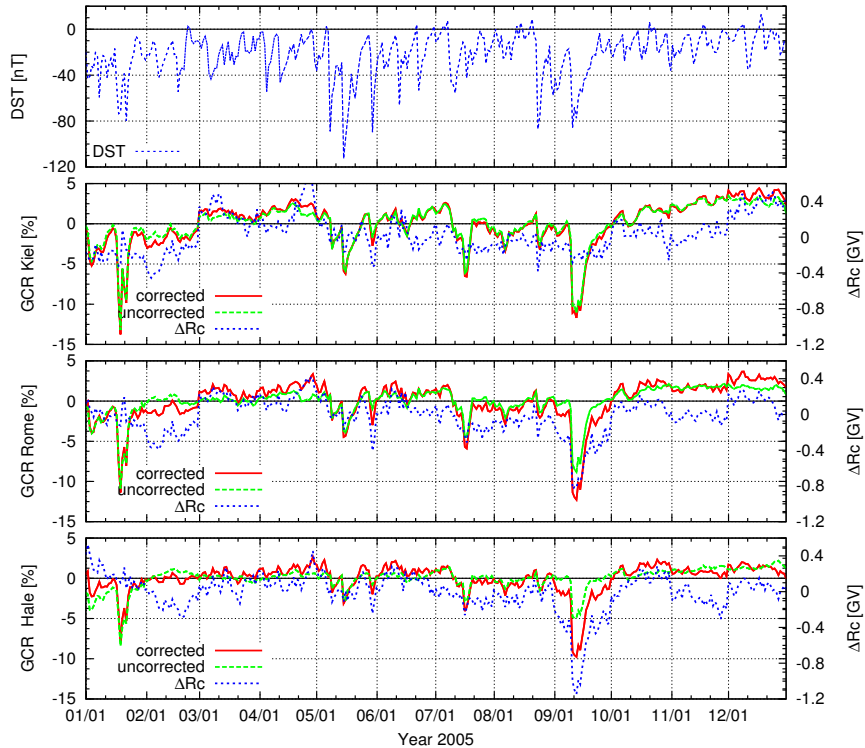}
\end{center}
\caption{\label{fig:fig1} Daily changes of DST index, cut-off variations $\Delta R_{c}$ and the corresponding GCR intensity for Kiel ($R_{c}=2.29$ GV), Rome ($R_{c}=6.32$ GV), and Halekala ($R_{c}=13.3$ GV) NMs corrected and uncorrected for geomagnetic disturbances for year 2005.}
\end{figure}
\begin{figure}[h]
  \begin{center}
\includegraphics[width=0.8\hsize]{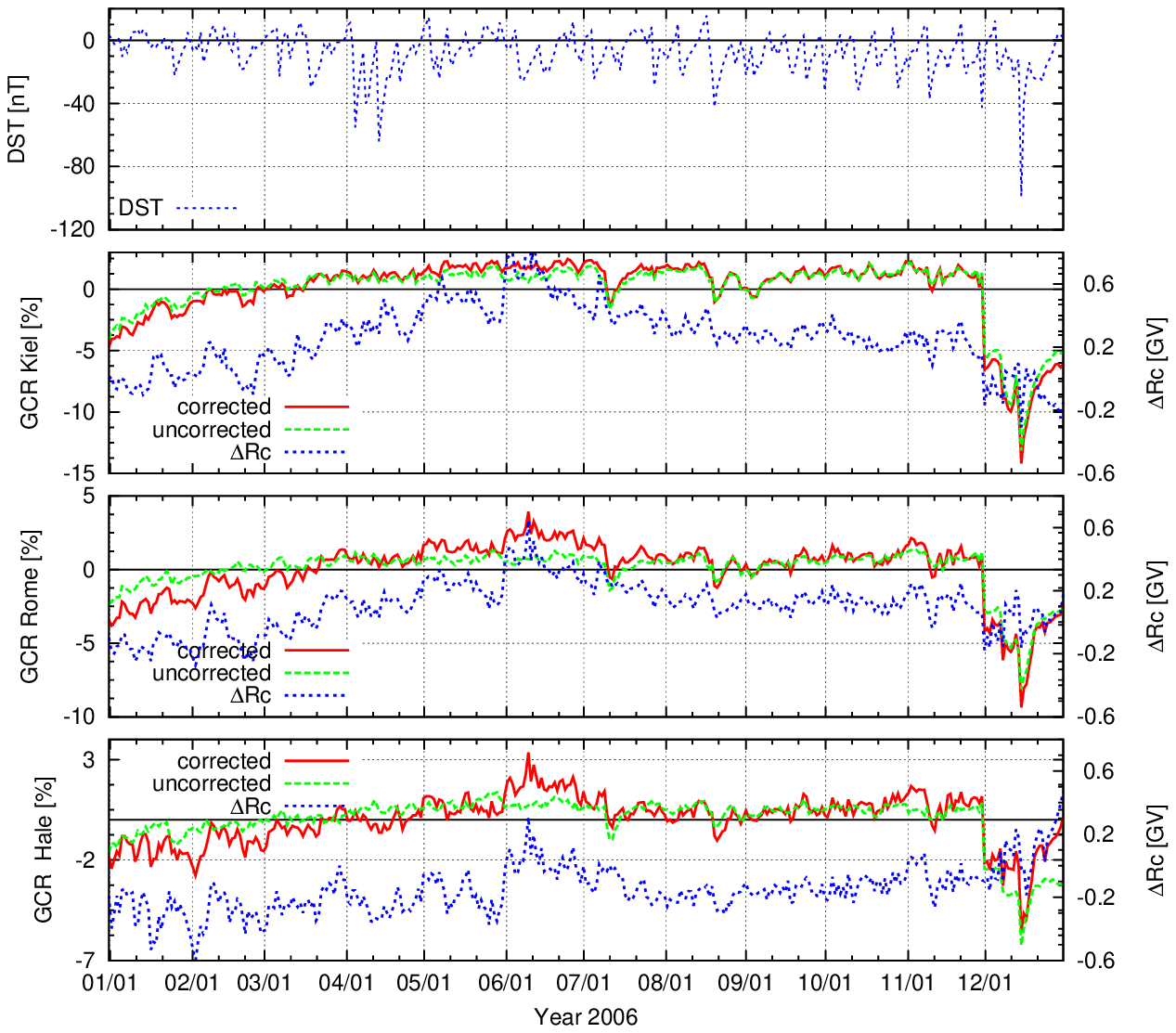}
\end{center}
\caption{\label{fig:fig2} The same as in Fig.~\ref{fig:fig1} but for year 2006.}
\end{figure}
\begin{figure}[h]
  \begin{center}
\includegraphics[width=0.8\hsize]{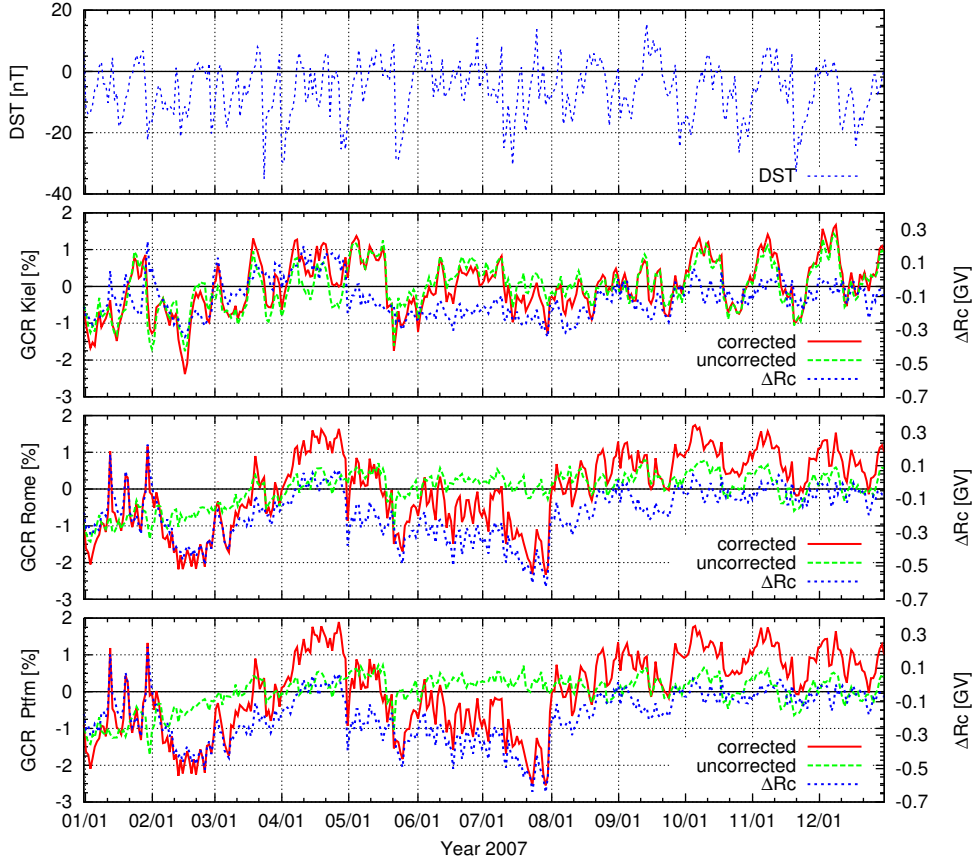}
\end{center}
\caption{\label{fig:fig3} The same as in Fig.~\ref{fig:fig1} but for year  2007. Due to shutdown of the Haleakala station the analogous changes for the Potchefstroom NM ($R_{c}=7.10$ GV) are presented.}
\end{figure}
\begin{figure}[h]
  \begin{center}
\includegraphics[width=0.8\hsize]{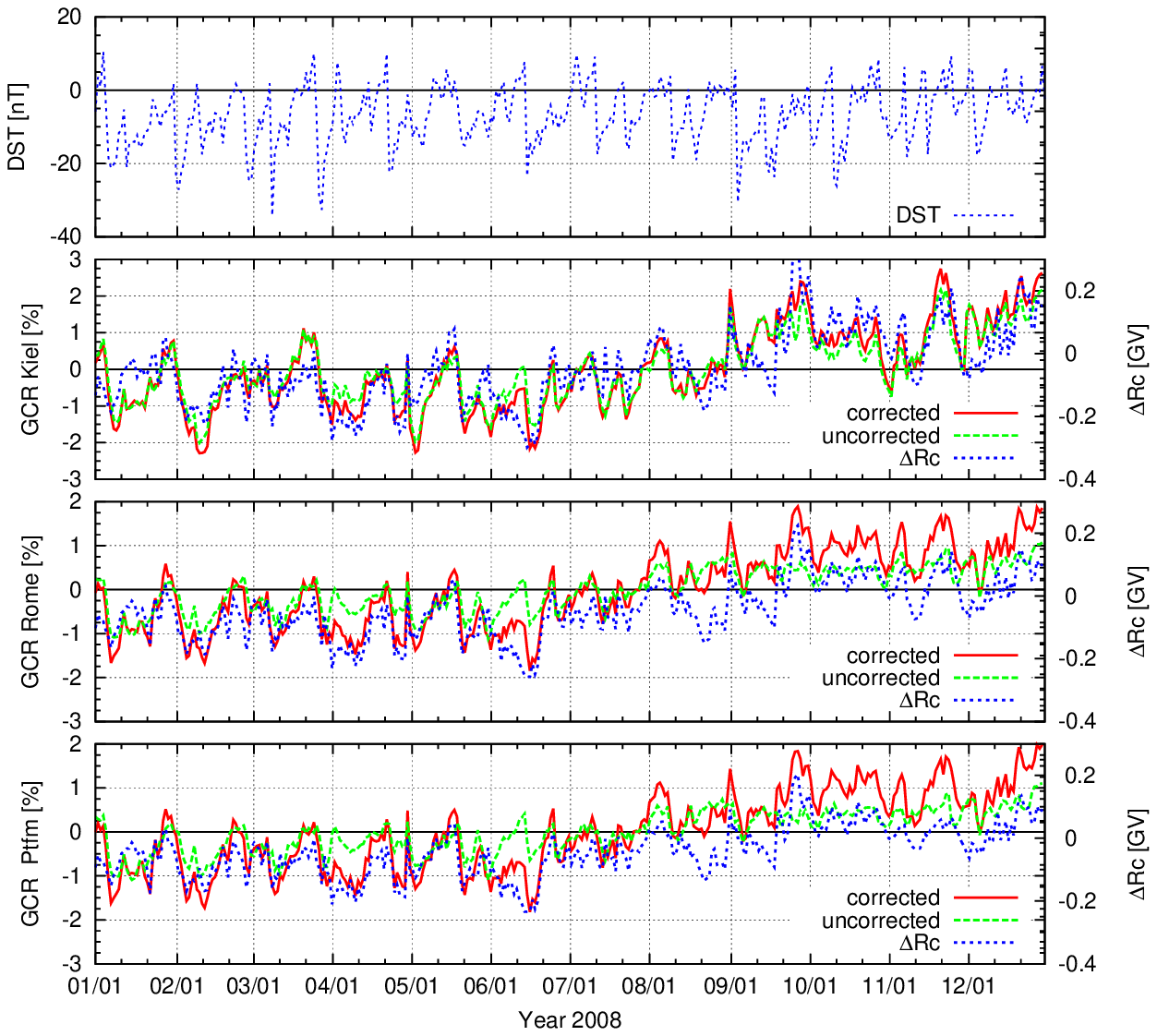}
\end{center}
\caption{\label{fig:fig4} The same as in Fig.~\ref{fig:fig3} but for year 2008. }
\end{figure}
\begin{figure}[!h]
  \begin{center}
\includegraphics[width=0.8\hsize]{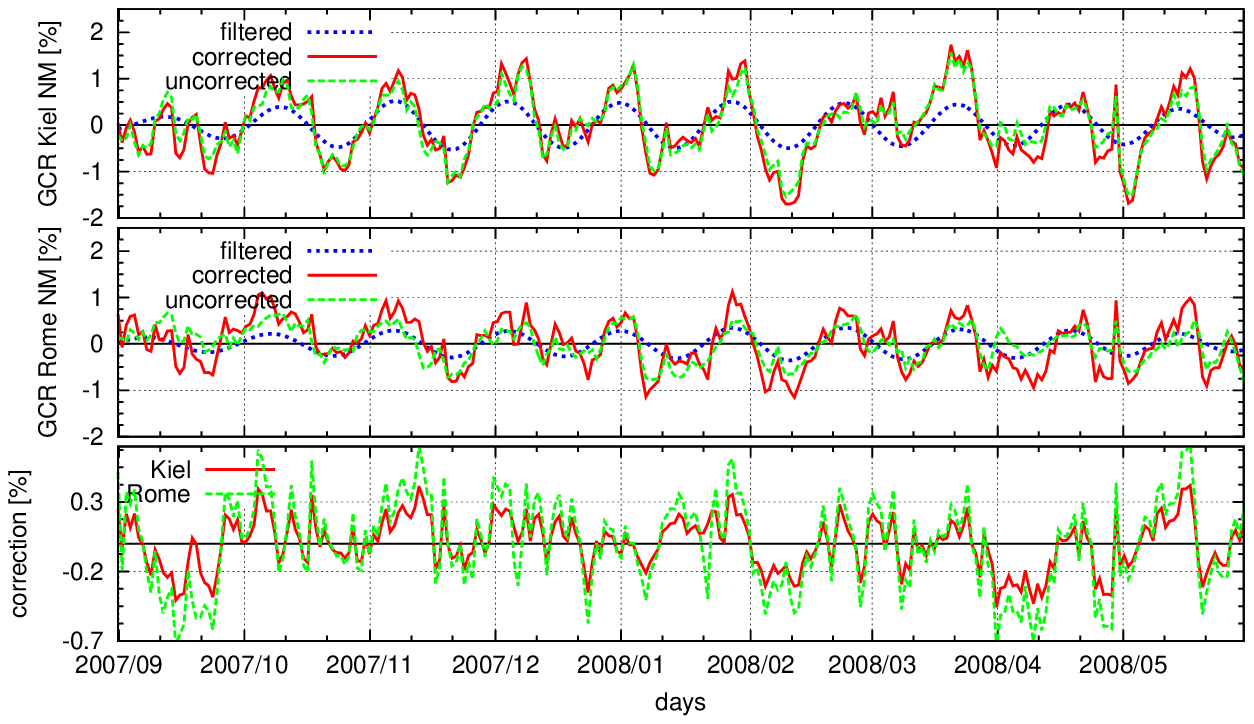}
\end{center}
\caption{\label{fig:fig5} Daily GCR intensity corrected and uncorrected for geomagnetic disturbances with filtered 27-day wave and the corresponding correction for Kiel and Rome NMs in 2007-2008.}
\end{figure}
\section{Introduction}
Magnetic cut-off rigidities for galactic cosmic ray (GCR) significantly changes depending on the level of disturbances of the Earth's magnetosphere. According to the Dungey mechanism \cite{Dungey} relatively high energy particles rush towards the Earth but are diverted around the Earth in circular orbits in the equatorial plane, forming a ring current at several earth radii, which causes significant geomagnetic field reduction. These reductions in the terrestrial magnetic field strength are measured by the Dst index (disturbance storm time index) \cite{Sugiura}.\\
\indent  An influence of short period  changes of the Earth's magnetic field connected with interaction of the Earth's magnetosphere and disturbed vicinity of the interplanetary space (coronal mass ejecta, shocks) on GCR intensity and anisotropy was  studied based on corrected and uncorrected neutron monitors (NMs) data in \cite{Tyasto,DS,AW1} and references therein.
At the same time,  assessment  of the 27-day variation of the GCR intensity by NMs data corrected for disturbances of Earth's magnetic field was not studied at all.  It is worth to underline that a survey of the 27-day variation of the GCR intensity is fundamental to understand  the influence of the heliolongitudinal asymmetry of solar wind and solar activity on behavior of the Earth's magnetosphere and atmosphere, and its consequences for the global climate processes. In order to address this issue an aim of  this paper is: (1) to estimate how the changes in the cut-off rigidities of various NMs influence the amplitudes of the 27-day variation of GCR intensity and (2) to study  changes in  rigidity spectrum of the 27-day variation of the GCR intensity in 2005-2008, the time interval of the unusually prolonged deep minimum  in the solar activity {\#}$23$.\\
\indent The recent solar minimum 23/24 provided  a unique opportunity to study recurrent variations of the GCR intensity under relatively stable conditions. Recurrent variations connected with corotating structures ($\sim27$ days), at the end of 2007 and almost for the whole year 2008  were clearly established in all solar wind and interplanetary parameters. Consequently, the 27-day recurrent variations of cosmic ray intensity were clearly  visible in a variety of cosmic ray counts of neutron monitors (e.g.,\cite{Alania10,MA13}) and space probes (e.g.,\cite{Leske13} ).

\section{Analysis of NMs data corrected and uncorrected for geomagnetic disturbances}
To find changes of cut-off rigidities $\Delta R_{c}$ for different neutron monitors during the analysed period of 2005-2008 we have  employed the spectrographic global survey (SGS) method developed by \cite{Dvornikov83}. Using ground-based measurements of GCR from the worldwide network of stations the SGS method allow to collect information on energy and pitch-angle distribution of primary GCRs in the interplanetary magnetic field (IMF), as well as on variations in the planetary system of geomagnetic cut-off rigidities per each observation hour. This method provides also the possibility to use the whole set of ground-based recording equipment (the worldwide network of neutron monitors at any level of the Earth's atmosphere, ground-based and underground meson telescopes, etc.) for analysis.  Along with phases of the first and second harmonics of the pitch-angle anisotropy, it can determine the rigidity spectrum of the isotropic component and anisotropy, obtain information about the IMF orientation from data on the phase of pitch-angle anisotropy. Moreover, the SGS method can determine variations in the GCR planetary system per each observation hour or per shorter time intervals when the CR intensity near Earth goes up during solar proton events. \\
\indent The essence of the SGS method is as follows: if amplitudes of variations in primary GCR intensity $\Delta J/J$  are small, then variations in intensity of secondary particles $\Delta I/I$  are connected with $\Delta J/J$  by the following relation:\\
\begin{eqnarray}\label{ii}
  \frac{\Delta I_{c}^{i}}{I_{c}^{i}}(h_{l})& = & -\int_{0}^{2\pi}\int_{0}^{\frac{\pi}{2}}\Delta R_{c}(\alpha,\beta)W_{c}^{i}[(R_{c}(\alpha,\beta),\alpha,\beta,h_{l}]sin\beta d\beta d\alpha+\\ \nonumber
& + &\int_{0}^{2\pi}\int_{0}^{\frac{\pi}{2}}\int_{R_{c}(\alpha,\beta)}^{\infty}\frac{\Delta J}{J}(R,\alpha,\beta)W_{c}^{i}(R,\alpha,\beta,h_{l})sin\beta d\beta d\alpha dR
\end{eqnarray}
where, $\frac{\Delta I_{c}^{i}}{I_{c}^{i}}(h_{l})$  is the amplitude of variations in flux of secondary particles of type $i$ observed at geographical point $c$ at height $h_{l}$ in the Earth's atmosphere at time $t$; $\frac{\Delta J}{J}(R,\alpha,\beta)$  are the variations in particle intensity at the boundary of the atmosphere at the given point; $\alpha$, $\beta$ are the azimuth and zenith angles of primary particles' arrival to the boundary of the atmosphere; $R$ is the magnetic rigidity of particles; $R_{c}(\alpha,\beta)$ is cut-off rigidity in directions $(\alpha,\beta)$, and $\Delta R_{c}(\alpha,\beta)$ are its possible variations. The $W_{c}^{i}(R,\alpha,\beta,h_{l})$ is the function determining a connection between  $\frac{\Delta I_{c}^{i}}{I_{c}^{i}}(h_{l})$ and $\frac{\Delta J}{J}(R,\alpha,\beta)$  satisfying the normalisation condition:
$\int_{0}^{2\pi}d\alpha \int_{0}^{\frac{\pi}{2}}sin\beta d\beta \int_{R_{c}(\alpha,\beta)}^{\infty}W_{c}^{i}(R,\alpha,\beta,h_{l}) dR=1$.
Assuming that direction of the particle motion in the given coordinate system is determined by the angles $\psi$ and $\lambda$, and the IMF direction by the angles $\psi_{0}$ and  $\lambda_{0}$, then the particle pitch angle $\theta$  can be expressed by following characteristics $\mu=cos\theta = sin \lambda sin \lambda_{0} + cos\lambda cos \lambda_{0} cos(\psi - \psi_{0})$.
Simultaneously, we can represent the function $\frac{\Delta J}{J}(R,\psi,\lambda)$  as a series
$\frac{\Delta J}{J}(R,\psi,\lambda)=\sum_{n=0}^{m}a_{n}(R)P_{n}(\mu)$
where $P_{n}(\mu)$ are the Legendre polynomials.
Taking into account  the simplifications and restrictions presented in detail in \cite{DS98} we obtain
\begin{equation}\label{ii2}
  \frac{\Delta I_{c}^{i}}{I_{c}^{i}}(h_{l})=-\Delta R_{c}W_{c}^{i}(R_{c},h_{l})+\sum_{j=0}^{10} \sum_{k=1}^{m_{n}}A_{jk}B_{jkc}^{il}
\end{equation}
where
$A_{0k}=a_{0k}$; $A_{1k}=a_{1k}cos\psi_{0}cos\lambda_{0}$,  $A_{2k}=a_{1k}sin\psi_{0}cos\lambda_{0}$; $A_{3k}=a_{1k}sin\lambda_{0} $,  $A_{4k}=a_{2k}cos^{2}\psi_{0}cos^{2}\lambda_{0}$, $ A_{5k}=a_{2k}cos\psi_{0}sin\psi_{0}cos^{2}\lambda_{0}$,  $A_{6k}=a_{2k}sin^{2}\psi_{0}cos^{2}\lambda_{0}$,
$A_{7k}=a_{2k}cos\psi_{0}cos\lambda_{0}sin\lambda_{0}$, $A_{8k}=a_{2k}sin\psi_{0}cos\lambda_{0}sin\lambda_{0}$, $A_{9k}=a_{2k}sin^{2}\lambda_{0}$,  $A_{10k}=a_{2k}$. The  coefficients  $B_{jkc}^{il}$ are determined based on the
asymptotic angles derived from calculations of the particles trajectory \cite{shea}.
Summarizing, the problem reduces to estimation of the following unknown parameters: $\Delta R_{c}$-showing the variation of cut-off rigidity; $a_{01}, a_{02}, a_{03}, a_{11}, a_{12}, a_{21}, a_{22}$ -showing amplitudes and rigidity dependencies of the zero, first and second harmonics of the pitch-angle GCR distribution; $\psi_{0}$ and  $\lambda_{0}$ - showing the IMF direction averaged over the Larmor circle (accurate to  $\pi$). To solve the problem in hand, it is essential that the network of stations and detector arrays at  stations used in the calculations provide abundance of the system of above mentioned equations  and linear independence of coefficients  $B_{jkc}^{il}$. There should be at least two detectors with different $W_{c}^{i}(R_{c},h_{l})$  at the stations where GCR variations are determinable. The details of the SGS method are presented in \cite{DS98}.\\
\indent Applying the SGS method we have calculated the daily changes of the $\Delta R_{c}$  by means of 27 neutron monitor stations hourly data during years 2005-2006 versus the cut-off rigidity $R_{c}$ on 10 Jun 2004, and for period of 2007-2008 versus $R_{c}$ on 5 November 2007.
Figs.~\ref{fig:fig1}-~\ref{fig:fig4}  present daily changes of Dst index, cut-off variations $\Delta R_{c}$ and the corresponding GCR intensity for Kiel, Rome, Potchefstroom and Haleakala NMs corrected and uncorrected for geomagnetic disturbances for 2005 (Fig.~\ref{fig:fig1}), 2006 (Fig.~\ref{fig:fig2}), 2007 (Fig.~\ref{fig:fig3}) and 2008 (Fig.~\ref{fig:fig4}).
Fig.~\ref{fig:fig1}-Fig.~\ref{fig:fig4} show that the corrected and uncorrected GCR intensity do not differ much for NMs with cut-off rigidities  smaller than 4-5 GV, while for NMs with higher cut-off rigidities the difference is as notable as cut-off rigidity increases i.e. a level of changes of  cut-off rigidities  is larger. One can see that the correction for geomagnetic disturbances is the largest when the rapid changes of Dst index are observed (related with shocks and Forbush decreases), but becomes weaker in quite stable conditions.\\
\indent In this paper we consider in detail 27-day wave in corrected and uncorrected GCR intensity and the effect of geomagnetic correction on periodic variations. Daily GCR intensity corrected and uncorrected for geomagnetic disturbances with filtered \cite{Otnes} 27-day wave  and the corresponding correction for Kiel and Rome NMs in 2007-2008 are presented in Fig.~\ref{fig:fig5}. As an example Fig.~\ref{fig:fig5} presents only few Bartel's rotations in 2007-2008 with excellent quasi-periodic changes ($\sim27$ days) in the GCR intensity.

\section{Rigidity spectrum of the 27-day variation based on the corrected and uncorrected NMs data}
Usually, a power law rigidity $R$ spectrum of any classes of the  GCR intensity variations is characterized by an exponent  $\gamma$, as $\sim R^{-\gamma}$. The exponent $ \gamma$ is an important instrument to study  a dependence of amplitudes of the GCR intensity changes  on  energy of primary GCR particles. The value of $\gamma$ is determined by the structure of the IMF turbulence responsible for the scattering of GCR particles to which neutron monitor respond (e.g.\cite{AIS,WA10,GA13}). A character of this dependencies  is different  for various classes of  the GCR intensity variations. In \cite{GA} was shown that temporal changes of the exponent $\gamma$ for the 11-year variations \cite{AIS} inversely correlates with the changes of the exponent $ \gamma$ for the 27-day variations of the GCR intensity.\\
\indent In this paper the values of the rigidity spectrum ($\delta D(R)/D(R)$) exponent $ \gamma$ of the 27-day variation of the GCR intensity were calculated  using both, corrected and uncorrected for geomagnetic disturbances data of  NMs with different cut-off rigidities.\\
\begin{figure}[h]
  \begin{center}
\includegraphics[width=0.7\hsize]{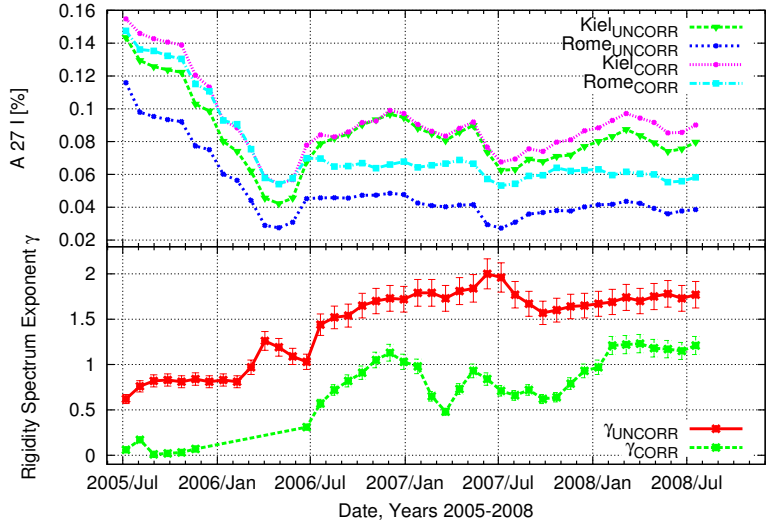}
\end{center}
\caption{\label{fig:fig6}(Top panel)The temporal changes in the amplitudes of the 27-day variation of the corrected and uncorrected GCR intensity for Kiel and Rome NMs; (Bottom panel) The exponent $ \gamma$ of the rigidity spectrum of the 27-day variation of the corrected and uncorrected GCR intensity  in 2005-2008.}

\end{figure}
We have calculated the amplitudes of the 27-day variation of the GCR intensity by means of daily data using the harmonic analyses method e.g.\cite{Gubbins2004} during each Bartel's rotation period:
\begin{equation}\label{e1}
  x(k\Delta t)=\frac{a_{0}}{2}+\sum ^{N/2}_{n=1} [ a_{n}cos(\frac{2\pi}{T}n k \Delta t)+b_{n}sin(\frac{2\pi}{T}n k \Delta t)],
\end{equation}
where $x(k\Delta t)$ designates the daily data of the GCR intensity, $N=27$ and the coefficients have a form:
$ a_{0}=\frac{2}{N} \sum ^{N}_{k=1} x(k\Delta t)$, $ a_{n}=\frac{2}{N} \sum ^{N}_{k=1} [ x(k\Delta t)cos \frac{2\pi k n}{N} ]$, $ b_{n}=\frac{2}{N} \sum ^{N}_{k=1} [ x(k\Delta t)sin \frac{2\pi k n}{N}]$.\\
Applying the above presented formulas we determine  the amplitude of the 27-day variation (A27I) of the GCR intensity as $A27I=\sqrt{a_{1}^{2}+b_{1}^{2}}$.\\
\indent The rigidity spectrum of the quasi-periodic variation was calculated by means of the amplitudes of the 27-day variation of the GCR intensity based on the method presented, e.g. in \cite{dorman63},\cite{GA}:
\begin{eqnarray}
\frac{\delta D(R)}{D(R)}= \left\{%
\begin{array}{ll}
    AR^{-\gamma} & \hbox{$R \leq R_{max}$}\\
    0 & \hbox{$R > R_{max}$}
\end{array}%
\right.
\end{eqnarray}
$R_{max}$ designates the upper limiting rigidity beyond which the
quasi-periodic variation of the galactic cosmic ray intensity
disappears (100 GV). Detailed description of the $ \gamma$  calculation is described in \cite{AW}.\\
\indent An investigation of the long-period changes of the power law rigidity spectrum of the 27-day variation of the GCR intensity (e.g. in the period 2005-2008) requires a data from neutron monitors with a long term stability. Unfortunately, only a few neutron monitors from worldwide network of stations satisfy this requirement. For this reason, we chose Kiel and Rome neutron monitors data to calculate the rigidity spectrum of  the amplitudes of the 27-day variation of the GCR in the period 2005-2008. To use two neutron monitors for the calculation of the reasonably reliable rigidity spectrum exponent $\gamma$ there must be considerable  difference between the cut-off magnetic rigidities ($R_{c}$). This demand is satisfactorily fulfilled for Kiel and Rome neutron monitors; for Kiel neutron monitor $R_{c}=2.29$ GV and for Rome neutron monitor $R_{c}=6.32$ GV (or the corresponding median rigidity of response for Kiel neutron monitor equals $17$ GV and for Rome - $23$ GV, respectively. A validity of this approach was shown in papers \cite{GA13, GA}.\\
\indent Fig.~\ref{fig:fig6} (top panel) demonstrates the temporal changes in the amplitudes of the 27-day variation of the corrected and uncorrected GCR intensity for Kiel and Rome NMs; and (bottom panel) the matching exponent $ \gamma$ of the rigidity spectrum of the 27-day variation of the corrected and uncorrected GCR intensity  in 2005-2008. Unfortunately, in the first half of the year 2006 calculation of the exponent $ \gamma$ based on the corrected data was not possible due to lack of the discrepancy between the amplitudes of the 27-day variation for the used NMs.
Fig.~\ref{fig:fig6} shows that rigidity spectrum of  the 27-day variation of the GCR intensity found by uncorrected data of NMs  is soft while it is hard  for corrected data. For both cases there is visible a clear tendency of the softening of the temporal changes of the rigidity spectrum from 2005 to 2008 (approaching minimum of solar activity) coinciding with results obtained in \cite{GA13}.

\section{Summary}
\begin{enumerate}
  \item Comparison of the amplitudes of the 27-day variation of the GCR intensity calculated based on the NMs data corrected and uncorrected for geomagnetic disturbances showed that these amplitudes do not differ for NMs with cut-off rigidities smaller than 4-5 GV in comparison with NMs of higher cut-off rigidities. This indicates that to study the 27-day variations of the GCR intensity by uncorrected data of NMs it is necessary to use data of NMs with cut of rigidities  smaller than 4-5 GV.
  \item Rigidity spectrum of the 27-day variation of the GCR intensity found by uncorrected data of NMs is soft, while it is hard for data corrected on geomagnetic disturbances. For both cases, there is an apparent tendency of softening temporal changes of the rigidity spectrum from 2005 to 2008 period approaching minimum of solar activity.
\end{enumerate}

\section*{Acknowledgments}
We are grateful to the Principal Investigators from worldwide network of neutron monitors and  OmniWeb database for access to the data.

\end{document}